# Phase Coherent Oscillations of Excitonic Photocapacitance and Bose-Einstein Condensation in Quantum Coupled 0D-2D Heterostructure

Amit Bhunia[1], Mohit Kumar Singh[1], Maryam Al Huwayz[2,3],

Mohamed Henini[2] and Shouvik Datta[1] *

[1]*Department of Physics & Center for Energy Science, Indian Institute of Science Education and Research, Pune 411008, Maharashtra, India*

[2]*School of Physics and Astronomy, University of Nottingham, Nottingham NG7 2RD, UK*

[3]*Physics Department, Faculty of science, Princess Nourah Bint Abdulrahman University, Riyadh, Saudi Arabia*

*Corresponding Author's Email: shouvik@iiserpune.ac.in

# ABSTRACT

We report quantum coherent oscillations of photocapacitance of a double-barrier resonant tunneling heterostructure with bias at 10 K. Periodic presence and absence of sharp excitonic transitions in photocapacitance spectra with increasing bias demonstrate strong coupling between InAs quantum dots (0D) and triangular GaAs quantum well (2D). Coherent resonant tunneling in this 0D-2D heterostructure establishes the momentum space narrowing of excitonic Bose-Einstein Condensation. Drastic increase of indirect exciton densities below 70 K reveal that excitonic wave functions anchored with each InAs quantum dots can laterally overlap across wide region around 200 μm to create a macroscopic quantum state of excitonic Bose-Einstein condensate. This itself points out the difficulties encountered in the usual 2D-2D bilayers and coupled quantum well samples used earlier to study excitonic BEC. Finally, we predict how coupled 'quantum-dots and quantum-well' heterostructures can display excitonic Bose-Einstein condensation at even higher temperatures.

**I. Introduction.**

Here we investigate how coherent resonant tunneling offers unique insights into many-body physics of spatially indirect bound state of electron-hole pairs or excitons, specifically about the 'elusive' experimental signatures of excitonic Bose-Einstein condensation (BEC). In general, semiconductor p-i-n heterostructures having zero dimensional (0D) quantum dots (QDs) were well studied in the contexts of both coherent Fabry-Perot type resonant tunneling and sequential resonant tunneling[1,2,3] of electrons. Earlier reports[4] of multiple peaks and/or oscillations in photocapacitance measurements were explained as charging of QD's S and P excitonic states[5] and correlations of excitonic complexes[6]. Recently, Vdovin et al. reported[7,8,9] interesting quantum oscillations in the laser induced photocurrent, photocapacitance measurements in a double barrier quantum heterostructure sample where InAs QDs are separated from two dimensional (2D) GaAs quantum wells by thin AlAs barriers. In this paper, we present further studies of quantum coherent photocapacitance oscillations and spectroscopy on one of these samples excited with an incoherent light source. Two sharp, resonant peaks in the photocapacitance spectra, visible only during peak resonant tunneling, clearly point towards the presence of quantum coupled indirect excitons in this 0D-2D heterostructure. We will explain how these phase coherent oscillations of excitonic quantum capacitance under light can be related to the presence of macroscopically large long-range order of BEC quantum state of spatially indirect, dipolar excitons.

Promising experimental observations of 'potential' Bose-Einstein condensation of excitons in 2D-2D bilayers of coupled quantum wells had already been reported[10,11] earlier including some more recent and interesting studies[12,13,14,15,16]. However, precise and unambiguous experimental signatures of both excitons and excitonic BEC were certainly complicated and non-trivial to interpret from many of these results. In fact, reports dealing with optical signature of excitonic BEC in coupled quantum wells were strongly debated[17,18] in the past. It is also reported that light emitting bright excitons in III-V semiconductors cannot be a part of the ensemble of excitons which forms the real Bose-Einstein condensate[19,20]. Most importantly, experimental evidences of excitons and signatures of the all-important off-diagonal long-range order as testimony to the presence of coherent macroscopic quantum state of excitonic BEC over a large region were rarely provided in the past. Here, we will argue that strong in-plane localization of holes within InAs QDs provide the overlapping bosonic quantum states necessary for these 0D-2D indirect excitons to undergo excitonic BEC over a wide region. We will explain how this further allows Fabry-Perot like coherent resonant tunneling through InAs QDs to produce the observed quantum coherent photocapacitance oscillations. In fact, excitonic BEC in these 0D-2D heterostructures necessitates strict restrictions on the in-plane momentum of 2D electrons which can take part in coherent resonant tunneling. Thereby, coherent resonant tunneling can act like an efficient probe to detect any momentum space narrowing associated with the excitonic BEC. It is also important to point out that this current work is a culmination of our previous investigations[21,22,23] to establish photocapacitance spectroscopy and capacitance spectroscopy as more powerful tools to detect experimental signatures of dielectric properties of dipolar, indirect excitons in quantum heterostructures. Most importantly, we will discuss how capacitive measurements can actually sense any emergent dipolar environment of excitons and associated enhancement of electric polarizations during the onset of excitonic BEC.

## II. Coupled 0D-2D 'quantum-dot & quantum-well' heterostructure and experimental observations of photo excited oscillations.

In Fig. 1a, a schematic diagram of the p-i-n type double barrier quantum heterostructure sample under reverse bias and selective photoexcitation is depicted above. Circular ring-shaped mesas having outer diameter ~200 μm are used for top electrical contact having an optical window for photocapacitance measurements. Below the sample structure, we also show the corresponding electronic band diagram. Electron-hole accumulations across the AlAs potential barriers are expected to increase with increasing reverse biases as both photo generated electrons in the p-type side and holes in the n-type side drift towards the AlAs barrier[22] in the steady state. This indicates the formation of GaAs triangular quantum wells (TQWs) for both electrons and holes, respectively, on opposite sides of AlAs barriers under photo excitation. Quantized energy levels for both electrons and holes inside InAs quantum dot are marked as $E_{QD}^e$ and $E_{QD}^h$, respectively[8]. As per earlier reports[8,24,25,26] on InAs/AlAs QDs, photo generated electrons can be in the quantized ground state energy level of InAs QDs, which are at $E_{QD}^e$ ~105 meV above the GaAs conduction band and photo generated holes at $E_{QD}^h$ which are ~15 meV above the GaAs valence band. We also want to emphasize here that formation of a two-dimensional electron gas (2DEG) inside GaAs TQW can happen only when the quasi Fermi level for accumulated electrons on the p type side of AlAs barrier crosses the GaAs conduction band edge in an ideal zero Kelvin picture. Similarly, formation of any two-dimensional hole gas (2DHG) inside GaAs TQW is possible when the quasi Fermi level for holes accumulated on the n type side of AlAs barrier crosses the respective GaAs valence band edge. However, it is expected that accumulation of photo generated holes in the form of 2DHG inside GaAs TQW will be much less under reverse bias because of lesser amounts of photo generations in the bottom n-type GaAs side of AlAs barrier as compared to that in the top p-type GaAs side of the AlAs barrier. Larger accumulations of electrons will also dictate that the 2DEG on the p-type side will have much stronger quantum confinement and have a deeper TQW as compared to the 2DHG on the n-type side of the AlAs barrier. On the other hand, we expect[22] a much larger hole accumulations in the p-type GaAs side under forward biases. Moreover, it must be noted that excess accumulations of either electrons or holes favor the formation of charged excitons or trions as also described previously[21].

In Fig. 1b, photocapacitance and photo conductance (G/ω) oscillations are plotted as a function of applied reverse bias at 10 K, where ω is the angular frequency of these impedance measurement across the top and bottom electrodes. Measurements were carried out at 10 kHz under selective photoexcitation of ~630±9 nm at a fixed light intensity of ~190 μW with a photon flux of ~3×10$^{16}$/cm$^2$ using a quartz-tungsten-halogen lamp. This broadband, non-coherent light source is particularly used to avoid any contributions from a coherent light source like a laser to generate these oscillations as well as not to forward bias the heterojunction with intense light absorptions. Further details of sample and experimental methods can be found below. Successive peaks and valleys of these oscillations are marked with respective bias values for later analyses. Clearly the relative phase between conductive and capacitive impedance also remains nearly constant with increasing reverse bias to produce these coherent, nearly sinusoidal oscillations. This can indicate a large scale phase matching of carriers over a wide region which is being probed using our photocapacitance measurements. Periodic oscillations of photo-G/ω are more pronounced with better signal to noise ratio than photocapacitance, although their relative magnitudes are nearly comparable. It is important to mention here that all these oscillatory changes of capacitance under photoexcitation are around a few hundreds of femto-Farads (fF) and these delicate coherent effects are possible to observe only when the current flowing perpendicular to the heterojunctions also remain small of the order of a few nano-amperes (nA). We also notice that these oscillations slowly die down with increasing currents under both biases. Therefore, the main aim of this work is to explore the roles of any emergent, cooperative condensed matter physics phenomena that can explain the robust, interferometric phase matching of charge carriers to generate these nearly sinusoidal oscillations measured using a large electrical top contact covering an area with diameter of ~200 μm.

In Fig. 1c, we plot photo $G/\omega$ and $dI_{Ph}/dV$ with respect to applied reverse bias, where $I_{Ph}$ is the DC photocurrent. Interestingly, these oscillations nearly superimpose with each other. Comparing Fig. 1b and Fig. 1c, we see that peaks of photocapacitance oscillations match exactly with positive peaks of photo-$G/\omega$, which also corresponds to positive $dI_{Ph}/dV$ or increasing $I_{Ph}$. Moreover, valleys of photocapacitance oscillations match exactly with valleys or negative peaks of photo-$G/\omega$, which further correspond to negative $dI_{Ph}/dV$ or decreasing $I_{Ph}$. We had reported earlier[22], how photocapacitance can actually sense the dipolar signature of photo generated, bias driven indirect excitons around the AlAs barrier and how photocurrents can also originate from photo generations of carriers throughout the bulk of those samples. So from Fig. 1c, it is likely that measured Photo-$G/\omega$ is similarly being affected by photocurrents coming from photo generated excitons everywhere in this sample. We will again come back to these issues while discussing spectroscopic results in Fig. 2.

In Fig. 1d, we plot photocapacitance oscillations for different photo excitation wavelengths under both reverse and forward biases to check if there are any particular optical spectroscopic transitions associated with these oscillations. This also serves as our control experiment to rule out any experimental artefact in these electrical measurements. Measured photocapacitance oscillation merges with flat dark capacitance even under selective photoexcitation for wavelength > 830nm (below the GaAs bandgap around 10 K). Therefore, it is clear that the observed oscillations survive as long as we photo generate enough electrons and holes in GaAs. Any unwanted contributions from electronic defects and interface states could have resulted in capacitance oscillations even under dark. It is also apparent that the frequency and amplitude of these photocapacitance oscillations are significantly enhanced in the forward directions. We will explain later why these observations are associated with coherent quantum oscillations between a macroscopic bosonic ensemble of excitons and a fermionic ensemble of negatively charged trions as we vary reverse biases. Similar quantum oscillations in magnetic systems are usually proportional to the area of Fermi surface in momentum space via Onsager rule[27]. This Fermi surface area increases with increasing effective mass. Therefore, we are likely witnessing a qualitatively similar electric field induced enhancement of oscillation frequency in forward bias due to larger accumulation of holes across AlAs barrier as compared to that in reverse bias. As such the tunneling probability and the tunneling current of holes are expected to be significantly smaller compared to electrons due to their heavier effective mass. However, as shown in Fig. 1d, we observe much larger oscillation amplitudes in forward biases due to larger accumulation of holes on the p type side of AlAs barrier and subsequent enhancement of the effective Fermi surface area. We will now discuss below how any emergent many-body quantum coherent state may spread over such a wide area to generate these oscillations and how these are being probed using our photocapacitance measurements. From now on, however, we will restrict ourselves to discuss observations made under reverse biases only.

## III. Why do we see two distinctly sharp, resonant excitonic transitions only during peak photocapacitance oscillations?

Here we want to investigate how the presence of excitonic states in photocapacitance oscillations can be established and how these excitonic states can change with photocapacitance oscillation under increasing reverse biases. We will also discuss the different excitonic states that can be probed using photocapacitance and photo-G/ω spectra. Both of these spectra are plotted under applied reverse biases (-0.40V, -0.66V, -0.92V) from successive peaks shown in Fig. 1b on the left panel as Fig. 2a. Photocapacitance and photo-G/ω spectra are also plotted under applied reverse biases (-0.53V, -0.80V, -1.07V) of successive valleys observed in Fig. 1b on the right panel as Fig. 2b. In Fig. 2a, we see two sharp, distinct, resonant peaks at ~1.52 eV and ~1.61 eV in the photocapacitance spectra. It is generally expected that these sharp spectral peaks are due to formation of excitons[21]. However, comparing Fig. 2a and Fig. 2b, we observe significant differences in excitonic spectral shapes between photocapacitance spectra taken either under peak oscillation biases or under valley oscillation biases. This is because photocapacitance spectra measured at valley biases of photocapacitance oscillation don't show the presence of these two distinctly sharp resonant peaks and we rather observe a broad spectral peak appearing in the middle. This striking spectral difference allows us to connect the presence or absence of excitons in Fig. 2 with photocapacitance oscillation shown in Fig. 1b. We also realize the significant influences of exciton formations in creating these photo induced oscillations under applied biases. However, both these two excitonic states peaking at energies ~1.52 eV and ~1.61 eV survive in photo G/ω spectra with positive photo-G/ω for peak biases and with negative photo-G/ω for valley biases. This is related to our above mentioned explanations[22] given along with Fig. 1c on how photo G/ω actually follows changes in photocurrents generated throughout the sample. Moreover, it is well known that if equal number of photo generated electrons ($n_{Ph}$) and holes ($p_{Ph}$) can drift away from the p-i-n junction, then these don't contribute to steady state photocapacitance which is $\propto (n_{Ph} - p_{Ph})$, however they do contribute to photocurrent which is $\propto (n_{Ph} + p_{Ph})$. These quantitative differences between photocapacitance spectra and photo-G/ω spectra will be clearer in the context of observed photoluminescence oscillation which will be described later.

In another control experiment shown in Fig. 3a, we do not see any similarly sharp resonant excitonic peaks in either photocapacitance spectra or photo-G/ω spectra measured at zero bias under similar photoexcitation intensities used earlier in Fig. 2. This fact in addition to results shown in Fig. 2b, clearly point out that direct photo generation in InAs QDs and in GaAs TQW are not that significant except under peak oscillation biases. It also indicates the critical role of applied electric field across the heterojunctions in observing these oscillations. Therefore, these results of photocapacitance spectroscopy do raise a few important questions on whether these sharp spectral peaks at ~1.52 eV and ~1.61 eV are actually not coming from photo generation within isolated stationary state energy levels of InAs QD and GaAs TQW respectively? Are these sharp spectral peaks observable only at peak oscillation biases related to any interferometric splitting of coupled quantum states of GaAs TQW and InAs QD in presence of significant resonant tunneling[28]? We will again address these key issues in the last paragraph of this section.

Using the band diagram shown in Fig. 1a, we now try to understand above results in terms of a few different possibilities of exciton formations using 630 nm selective photo excitations in this structure under reverse bias only. The ~1.61 eV spectral peak is nominally expected to originate from strong photo excitations of direct excitons within InAs QDs as reported specifically for InAs/AlAs heterostructures[24,29,30]. However, as already mentioned above, we neither see this peak under zero applied bias as in Fig. 3a, nor we see it under valley oscillation biases as in Fig. 2b. On the other hand, the ~1.52 eV spectral peak is similar to the well-known excitonic transition of GaAs based heterostructures at these low temperatures[25]. Therefore, we think this 1.52 eV excitonic peak may originate from photo generation of electrons inside $E_{TQW}^e$ near the top p-type side of GaAs TQW and holes in – (a) either inside $E_{TQW}^h$ of bottom n-type side of GaAs TQW or (b) in $E_{QD}^h$ of InAs QDs. Both of these options correspond to formation of spatially indirect excitons[12,13]. However, more strongly bound indirect excitons can form with holes inside these InAs QDs. Later we will also argue how our estimates based on exciton Bohr radius from quantum capacitance mostly matches with the dipolar sizes of these InAs-GaAs indirect excitons only.

Due to the presence of AlAs barriers, there is enhanced accumulation of photo generated carriers within the GaAs TQW with increasing reverse bias. This increasing accumulation increases the applied electric field across the GaAs TQW. Quantized energy levels of electrons in this GaAs TQW can shift with applied bias or electric field. The n$^{th}$ energy level of GaAs TQW having infinite barriers for electrons can be written as[7]

$$E_n \propto \left[\frac{3\pi}{2}\left(n - \frac{1}{4}\right)\right]^{2/3} \left[\frac{(|\vec{F}_z|e\hbar)^2}{2m^*}\right]^{1/3} \tag{1}$$

where $E_n$ is the n$^{th}$ energy level of the TQW and $V(z) = e\vec{F}_z \cdot \vec{z}$ is the potential drop across the intrinsic region of the heterojunction, $\hbar$ is the reduced Planck constant, $|\vec{F}_z|$ is amount of electric field along the growth direction $\hat{z}$, $m^*$ is effective mass of electrons in GaAs. As a result, the quantized ground state energy level of $E_{TQW}^e$ also blue shifts with increasing electron accumulations under increasing reverse biases. Such blue shifts will be lesser for InAs QD levels due to its smaller width along the growth direction. As a result, the spectral blue shift (~17 meV) of indirect exciton related photocapacitance peak around 1.52 eV is nearly twice than that of ~9 meV blue shift of the InAs exciton related photocapacitance peak around 1.61 eV. Table. 1 shows that successive blue shift of the indirect exciton peak as the bias gradually increases from -0.40 V to -0.66 V to -0.92 V, can be explained using the above formula for first quantized energy level of the ground state of electron TQW. Moreover, the observed blue shifts of both peaks with increasing biases are also indicative of stronger coupling between InAs QDs and the 2DEG in GaAs TQW.

To summarize, we had already demonstrated in Fig. 1d, that photocapacitance oscillations survive even when selective photoexcitation wavelength remains much below the ~1.61 eV excitation for photoexcitation inside InAs QDs. Therefore, any direct 1.61 eV photoexcitation within InAs QD is certainly not necessary to generate these oscillations. Besides, any resonant tunneling between GaAs TQW and InAs QDs will always require matching of energy levels. However, it is not at all clear whether we observe any simple-minded matching of energy levels of $E_{TQW}^e$ of GaAs and $E_{QD}^e$ of InAs from the photocapacitance spectroscopy shown in Fig. 2. These spectral observations may actually indicate the presence of ~100 meV splitting[28] of energy levels of coupled quantum states of 0D-2D heterostructure once photo generated electrons take part in sizable resonant coherent tunneling at peak oscillation biases[28,31,32,33]. This likely also generate these sharply defined quantum states which can then be excited with ~1.52 eV and ~1.61 eV photons. This brings us to sort out the set of other relevant issues which will be discussed in the next two sections before we explain these spectroscopic observations any further.

**IV. Incoherent resonant tunneling vs coherent, phase conserving resonant tunneling.**

We now try to understand the physical mechanism of these phase matched sinusoidal oscillations over a wide area below our electrical mesa contact. This will require us to understand whether the resonant electron tunneling across the AlAs barrier between 2D GaAs TQW through discrete energy level of 0D InAs QDs is either coherent or incoherent. Here we assume from the beginning that tunneling of holes will be less significant due to their increased effective mass and also due to a smaller number of holes accumulated around AlAs under reverse bias. However, it is expected that in the steady state, some amount of holes can be there inside InAs QDs under finite biases. Ordinarily, such oscillations in photocapacitance may actually come from a few simple possibilities of how the density of these indirect excitonic dipoles can change with increasing bias. These can happen via three different ways – (i) sequential electron emission from successive bound states of 2DEG in the GaAs TQW towards the p-type side and (ii) periodic emptying of the successive quantum ground states of GaAs TQW due to non-coherent resonant electron tunneling from GaAs 2DEG to InAs QDs with increasing reverse bias. However, it is unlikely that electron Fermi levels will be situated at the top edge of the GaAs TQW barrier for this 2DEG, which can contribute to successive electron escape towards the p type side with increasing reverse bias. Moreover, we expect more electron and hole accumulations near AlAs barriers and subsequently a deeper TQWs with increasing biases. A deeper TQW and successive bound to unbound transition of quantized TQW levels with electron escape from the higher levels to top p-type side would have resulted in decreasing bias gaps of successive oscillation peaks with increasing biases, which is not observed in Fig. 1. Therefore, we rule out the first possibility mentioned above. Moreover, one can approximate $E_n$ of this GaAs TQW for $n = 6$ as 0.37 eV. One can compare this with the Γ-X conduction band discontinuity of $\Delta E_C$ ~0.42 eV for the GaAs/AlAs heterojunction using the usual 60/40 rule for estimating band discontinuities. So, using Schrodinger-Poisson type calculations one can easily estimate the number of quantized electron energy levels within GaAs TQWs which can be at most ~10 in numbers. However, the observed numbers[7] of finite oscillation peaks are much more than that and clearly cannot be explained by periodic emptying of TQW ground state with increasing biases. Consequently, we rule out the 2nd possibility too. The 3rd possibility of (iii) repeated, non-coherent, sequential resonant tunneling between the 1st TQW level and 1st InAs QD level would have also shown progressive reduction of the observed periodicity of photocapacitance/photocurrent oscillations. This is because we expect narrower TQW with higher quantum ground state energy level with increasing charge accumulation at higher biases. However, we don't observe this either. Besides, the discrete quantum ground state energy levels of InAs QDs and GaAs TQW and corresponding signatures of resonant tunneling need not suddenly vanish above 70 K. Additionally, it had already been reported[7] that the frequency of photocurrent oscillations increases with decreasing photo excitation intensity. However, the allowed numbers of quantized levels in GaAs TQW decrease with decreasing electron accumulation at lower photo excitation intensities. Therefore, these photocapacitance oscillations can never be explained with any simple-minded energy level matching scheme for sequential, incoherent tunneling processes. This is also because, all charge carriers participating in resonant tunneling must have a unique phase for all InAs QDs to produce these seemingly 'single phase' oscillation. Any random differences of phases among the charge carriers during resonant tunneling through many InAs QDs spread around 200 μm wide region could have easily averaged out such oscillations. Therefore, we need to explore other possible explanations for the observed periodic oscillations beyond these simple-minded incoherent tunneling scenarios. We will later discuss why precise momentum matching will be required and then explain the origin of coherent electron tunneling between GaAs TQW and InAs QDs.

## V. Observation of photoluminescence oscillations and negative photoconductivity for Trions.

Interestingly, from Fig. 2a we already found that for biases corresponding to peak photocapacitance oscillations, we observe the presence of both ~1.52 eV and ~1.61 eV resonant transitions of excitons in the photocapacitance spectra. However, at applied biases corresponding to valleys of photocapacitance oscillations, these two sharp spectral transitions are vanished. To understand these, we plot Fig. 3b. Here we see that the oscillations of DC-photocurrent match well with oscillation in the net light emission intensity using a front-collection photoluminescence (PL) geometry. We also understand why increasing positive $dI_{Ph}/dV$ can lead to more electrons being transferred from the top p-type side to bottom n-type side of the barrier junction under reverse bias as shown in Fig. 1c. Therefore, during positive Photo-G/ω, these electrons can actually recombine with holes accumulated either in the GaAs TQW located at the bottom n-type side of GaAs or even inside InAs QDs. On the other hand, during negative Photo-G/ω, these electrons can actually tunnel back to GaAs 2DEG from InAs QDs. This explains the oscillation of net PL shown in Fig. 3b as well as why the spectroscopic signatures of excitonic photo generation in photo-G/ω spectra can still remain visible at biases corresponding to oscillation valleys as in Fig. 2b. This is because photo-G/ω merely reflects the variation of $dI_{Ph}/dV$ due to changes in tunneling conductance at such biases. Any holes, photo generated in the GaAs TQW, are also expected to quickly drift away to produce $I_{Ph}$ or generate photoluminescence in presence of excess electrons. However, any photo generated electrons or holes which are not accumulated around the double barrier heterojunction cannot affect the steady state photocapacitance measured at 10 kHz. These conclusions are also supported by earlier[22] observations on why photocapacitance spectra can sense the changes in the dipolar environments of indirect excitonic dipoles only at the junction but photocurrents spectra can be affected by photo generation of excitons throughout the sample.

Moreover, we always expected more electrons to accumulate around p-type GaAs side of AlAs barrier than holes in the n-type side of the AlAs barrier. Therefore, we can't rule out formations of negatively charged trions around AlAs potential barrier due to the presence of excess electrons during one half of the oscillation process. Additionally, we have noticed in connection with Fig. 1b and Fig 1c, that peaks of photocapacitance oscillations match exactly with positive peaks of photo-G/ω, which also corresponds to positive $dI_{Ph}/dV$ or increasing $I_{Ph}$. Therefore, it is likely that the presence/absence of tunneling maxima/minima may decide whether we are forming a coupled bosonic system of direct and indirect excitons or a predominantly fermionic system of indirect trions across the AlAs barrier. It was also reported that the formation of trions lead to negative[34] photoconductivity due to strong many-body interactions. It is already evident from Fig. 1c that formation of predominantly fermionic system of negatively charged indirect trions at valley oscillation biases also coincide with negative peaks of photo-G/ω, which corresponds to negative $dI_{Ph}/dV$ or decreasing $I_{Ph}$. The broad photocapacitance peak at valley oscillation biases corresponds to minima of photocapacitance as well as to negative photo-G/ω. Therefore, the broad photocapacitance spectra in Fig. 2b are likely associated with formation of mixed dipolar ensembles of charged excitonic complexes including negatively charged trions, when electrons are tunneling back towards GaAs TQW at the tunneling minima. At this stage, it is beyond the scope of this study to investigate the precise many-body physics mechanisms which lead to this extensive broadening of the photocapacitance spectra at reverse biases corresponding to oscillation valleys as shown in Fig. 2b.

**VI. Observation of Photo excited 'quantum capacitance' and corresponding identification of indirect excitonic dipoles.**

We further investigated the role of quantum capacitance for this coupled 0D-2D excitonic systems in the neighborhood of a strong 2DEG inside the GaAs TQW. We expect that the presence of this strong 2DEG can substantially screen the excitonic charges and modify the total effective capacitance of the system under reverse bias. Quantum capacitance $(C_q)$ of such a structure is given by[35]

$$C_q = \rho e^2 = \frac{m^* e^2}{\pi \hbar^2} \qquad (2)$$

where $\rho$ is the two-dimensional density of states of the GaAs TQW, $C_q$ is quantum capacitance per unit area, $m^*$ is the effective mass, $e$ is electronic charge, $\hbar$ is the reduced Planck's constant. Now, for a series configuration of geometric capacitance ($C_g$) and $C_q$, one can write the effective capacitance ($C_{eff}$) as[36,37]

$$\frac{1}{C_{eff}} = \frac{1}{C_g} + \frac{1}{C_q} \tag{3}$$

Assuming $C_{eff}$ = 13 pF as the measured photocapacitance of the 1.52 eV spectral peak and $C_g$ = 12.8 pF as dark capacitance at -0.4 V reverse bias, we get $C_q$ = -832 pF. One can also rewrite this expression as[35]

$$C_q = \frac{m^*e^2}{\pi\hbar^2} = \frac{\varepsilon}{a_0^*/4} \tag{4}$$

$$a_0^* \equiv \frac{4\pi\varepsilon\hbar^2}{m^*q^2} = (0.053)\frac{m_0}{m^*}\frac{\varepsilon}{\varepsilon_0} \tag{5}$$

where $a_0^*$ is the effective Bohr radius in nm. One can understand[35] this $C_q$ as the effective quantum capacitance of a parallel plate capacitor with plates (dipoles) separated by a distance of $a_0^*/4$. Now using the modified expression of quantum capacitance given in Eq. 4, one can estimate $a_0^*$ as ~18 to 30 nm. Here we assume the static dielectric constant for GaAs as $(\varepsilon/\varepsilon_0)_{GaAs} \cong 12.5$. Accordingly, the distance between the plates of this equivalent, quantum parallel plate capacitor is $a_0^*/4$ ~5 nm. This estimate matches well with the minimum expected sizes of these 0D-2D indirect excitonic dipoles across a 5 nm AlAs barrier in between InAs QDs and GaAs TQW.

Therefore, the coherent oscillations observed in photocapacitance can be associated with phase matched, cooperative oscillations of the effective parallel plate capacitor of these quantum coupled 0D-2D indirect dipolar excitons. Such oscillations of quantum photo capacitance and effective Bohr radius ($a_0^*$) under reverse bias are shown in Fig. 4a. However, we note that similar explanations of photocapacitance under forward bias in presence of 2DHG for excess holes accumulated on the p-type side of AlAs barrier are quantitatively more complicated and possibly not feasible in terms of quantum capacitance as shown in Fig. 4b. One expects increased electron accumulations in GaAs TQW at increasing reverse biases as well as increasing tunneling currents. This decreases the quantum capacitance ($C_q$) and subsequently increases the dipolar separation ($a_0^*/4$) of these indirect excitons according to Eq. 4. We had already observed and reported[22] similar reduction of photocapacitance after the onset of tunneling. This interesting analogy of quantum capacitance with excitons was, however, not reported in the past. We must also add that observation of excitonic quantum capacitance and its oscillations do not preclude coherent resonant tunneling.

## VII. Sudden enhancement of electric polarization around 70 K and the onset of coherent electron tunneling between GaAs TQW and InAs QDs.

In Figs. 5a and 5b, we plot temperature dependent oscillations of photocapacitance and photo-G/ω, respectively. It is clear that oscillation magnitudes suddenly increase with decreasing temperature below 70 K. We will now discuss how this uncharacteristic increase in the measured photocapacitance is reminiscent of some kind of phase transition within this system of indirect excitonic dipoles. Does it also indicate any underlying quantum cooperative effects of these excitons over such a large area? To understand, we plot the variation of measured photocapacitance and photo-G/ω magnitudes for -0.4 V reverse bias in Fig. 5c. Although the magnitude of oscillation of photo-G/ω is only finite below 70 K but the variation of photocapacitance with temperature is clearly non-monotonic. Starting from room temperature, the measured photocapacitance decreases with lowering temperature as expected for any normal junction capacitance under thermally activated charge trappings in presence of electronic defects.

However, we see a sudden unexpected increase in photocapacitance magnitude below 70 K, which also coincides with the onset of coherent photocapacitance oscillations. This sudden increase at lower temperatures cannot be explained in terms of usual freezing of thermally activated shallow defect states. As such, surface charge density of photo generated carriers ($\sigma_{ph}$) per unit area around the AlAs/InAs/AlAs heterojunction can be estimated from the following equation[21]

$$\sigma_{ph} e = CV(z) \qquad (6)$$

where $e$ is the electric charge and $C$ is the peak value of excitonic photocapacitance per unit area at each bias ($V$) applied along the z direction. One can also relate this $\sigma_{ph}$ to the density of dipolar charges of these indirect excitons per unit area in a standard parallel plate capacitor configuration as[21]

$$\sigma_{Ph} = \langle \vec{P} \rangle \cdot \hat{z} \qquad (7)$$

where $\langle \vec{P} \rangle$ is the average electric polarization vector of these indirect excitonic dipoles and $\hat{z}$ is the unit vector along the growth direction of this sample. Measured photocapacitance is, therefore, proportional[21] to macroscopic orientational average of the electric polarization vector $\langle \vec{P} \rangle$ of all indirect excitonic dipoles below the ring shaped top mesa contact having outer diameter ~200 μm. Therefore, this sudden increase of photocapacitance indicates a drastic enhancement of $\langle \vec{P} \rangle$ as well as an abrupt increase of the number density of these indirect excitons or bosonic dipoles in terms of $\sigma_{Ph}$. This can only happen if dipole moments of these indirect excitons undergo some sort of density driven phase transition around 70 K and begin to align themselves along $\hat{z}$. Consequently, the surface charge density corresponding to electric polarization of these excitonic dipoles increase from $\sigma_{Ph} \sim 1.067 \times 10^{11}/cm^2$ at 70 K to $\sigma_{Ph} \sim 1.099 \times 10^{11}/cm^2$ at 10.5 K, which is estimated from the data plotted in Fig. 5c. Before we start to discuss the cause for such density driven phase transition of these excitonic bosons, physical understanding of restricting the in-plane momentum of electrons in GaAs TQW will now be explained first.

This double barrier quantum heterostructure can act like a Fabry-Perot interferometer[38] for electrons tunneling through AlAs barriers. So coherent resonant tunneling through InAs QDs in the presence of negligible scattering are expected[38] to demonstrate such damped, nearly single-phase sinusoidal oscillations of both photo generated tunneling current which is proportional to tunneling transmission probability as well as photocapacitance which is proportional to density of state of these dipolar surface charge or $\sigma_{Ph}$. Coherent interference[38] of tunneling currents is only possible if most carriers (here electrons) can traverse the ~1 nm height of all these InAs QDs without significant loss of phase randomization and get reflected from the AlAs barriers repeatedly in the process. The most striking observation that stands out here in our experiments is the fact that such coherent phase matching during tunneling of all constituent electrons of these indirect excitons has to take place throughout the ~200 μm wide area in the XY plane of the GaAs TQW below the top electrical contact. This leads us to ask - does this indicate any underlying quantum cooperative effects over a large region?

As mentioned above, we assume that electrons are predominantly taking part in tunneling across AlAs barrier due to their smaller effective mass. Interestingly, coherent resonant tunneling between electrons of the 2DEG and zero dimensional InAs QDs also requires stringent energy-momentum conservation for exact Fermi level nesting. This is because free electrons in 2DEG can possess large transverse momentum in the plane of the 2DEG. Therefore, electrons of this 2DEG can tunnel coherently to an ensemble of InAs QDs, only and only if their in-plane momentums $\vec{k}_x$, $\vec{k}_y$ are such that the resonant tunneling condition is satisfied for the ground state quantized level ($E^e_{QD}$) of electrons in all of these InAs QDs. The energy conservation requires[39]

$$E^e_{QD} = E^e_{TQW} + \frac{\hbar^2}{2m^*}(\vec{k}_x^2 + \vec{k}_y^2) + (eV(z)) \tag{8}$$

where $E^e_{QD}, E^e_{TQW}$ are ground state quantized energy level of electrons in InAs QD and in GaAs 2DEG, respectively, $e$ is the electronic charge and $V(z)$ is the applied bias across the heterojunction along the $\hat{z}$ axis. Coherent resonant tunneling between quantum levels of 2DEG and all these QDs at a particular, constant bias, say $V(z) = V_0$, is possible only if the magnitude of transverse momentum $|\vec{k}_e^{2DEG}(V_0)| = (k_x^2(V_0) + k_y^2(V_0))^{1/2}$ of all electrons within the wide XY plane of GaAs 2DEG, become nearly identical.

Under any substantial resonant tunneling, coupled quantum states of InAs QDs and GaAs 2DEG will make electrons photo generated either in InAs QD or in GaAs TQW to remain quantum mechanically indistinguishable[28]. In addition, presence of momentum quantization of holes in zero-dimensional states within InAs QDs having $\vec{k}_h^{InAs}(V_0)$ can also restrict electrons in GaAs TQW having an unique $\vec{k}_e^{2DEG}(V_0)$. This is because the net center of mass momentum $(\vec{K}_{Exciton}(V_0))$ of these indirect excitons within coupled quantum states of InAs QDs and GaAs 2DEG must satisfy

$$\vec{K}_{Exciton}(V_0) = \vec{k}_e^{2DEG}(V_0) + \vec{k}_h^{InAs}(V_0) \cong 0 \tag{9}$$

Therefore, $\vec{k}_e^{2DEG}$ needs to approach a unique quantized value as $\vec{k}_e^{2DEG}(V_0) \approx -\vec{k}_h^{InAs}(V_0)$ for any resonant photo generations of excitons with $\vec{K}_{Exciton}(V_0) \cong 0$. This can happen, firstly, by absorbing resonant photons having comparatively negligible momentum than electrons, once these coupled quantum states with respective transition energies of ~1.52 eV and ~1.61 eV are created in the steady state. Secondly, by the fact that these coupled 0D-2D indirect excitons are not free in the XY plane but remain spatially anchored with individual InAs QDs.

Electrons with unique $\vec{k}_e^{2DEG}(V_0) = \vec{k}_e$ can then easily take part in coherent resonant tunneling process within this double barrier AlAs/InAs/AlAs Fabry-Perot tunneling structure and repeatedly reflect back and forth coherently. Assume that the vertical height of these InAs QDs along growth direction are taken as $\vec{z}_{InAs}$. Therefore, a coherent buildup of one particular phase $(\varphi = \vec{k}_e . \vec{z}_{InAs})$ of these electrons inside this resonant Fabry-Perot interferometer will be required for any coherent resonant transmission through all these InAs QDs which are spread over a large area with diameter ~200 μm. We understand that such long-range cooperative phenomena with a unique phase within this Fabry-Perot resonator structure can subsequently give rise to phase matched photocapacitance and photo-G/ω oscillations as well as two sharp spectral signatures of excitons only at the peak oscillation biases as shown in Fig. 2a. This certainly necessitates long-range phase cooperation during resonant tunneling in these coupled quantum structures. Presence of any unwanted phase non-conserving, incoherent and inelastic thermal processes can easily destroy this Fabry-Perot interference effects spanning over a ~200 μm wide area. We indeed see that observed photocapacitance oscillation actually vanishes at temperatures above 70 K as well as slowly decays at higher biases. Therefore, increase of oscillatory photocapacitance at temperatures below 70 K is a clear indication of not only coherent resonant tunneling, but also to a density driven phase transition to a coherent many-body excitonic state with constant $\vec{K}_{Exciton}(V_0) = 0$ as its main driving mechanism. Therefore, observation of coherent oscillation due to bias induced resonant tunneling in this coupled 0D-2D heterostructure may be acting like a precise 'momentum space filter' through which the experimental signature of excitonic BEC in terms of momentum space narrowing is actually being detected here. Detailed implication of this understanding will be discussed in the next section where we will explain how this unique choice of $\vec{k}_e^{2DEG}(V_0)$ necessary for coherent resonant tunneling originates from underlying BEC of these indirect excitons.

**VIII. Why do we expect a macroscopically large quantum state of excitonic BEC?**

In order to understand how the above mentioned perfect matching of transverse momentum of electrons in the XY plane of the 2DEG may actually happen during resonant tunneling, we look closely at the formation of spatially indirect excitons under reverse bias. This is even more evident from the two sharp, resonant excitonic peaks of photocapacitance and photo-G/ω spectra in Fig. 2a. Strong coulomb attraction between electrons and holes of these indirect excitons in this coupled 0D-2D system can be considered as an ensemble of bosons which are strongly localized in transverse XY plane due to immobile InAs QDs. We now explore the possibility of BEC of these indirect excitons populating a macroscopically coherent quantum state with singularly unique and large $\langle\vec{P}\rangle$ below 70 K. An estimate of the critical[40] surface density of 2D bosons as the order parameter for BEC is given as

$$N_{BEC}^{2D} = 2\left(\frac{1}{\lambda_{Th}}\right)^2 ln\left(\frac{L}{\lambda_{Th}}\right) \tag{10}$$

where $\lambda_{Th}$ is the thermal de-Broglie wavelength for these excitons, L being the lateral extent of the 2D system = the diameter of our top electrical contact ~200 μm. Using $m^*$ for excitons in GaAs as $0.058m_0$ where $m_0$ is the free electron mass, we assume that the thermal wavelength for these indirect excitons can be[40]

$$\lambda_{Th} = \left(\frac{2\pi\hbar^2}{m^* k_B T}\right)^{1/2} \approx 100 \text{ nm} \tag{11}$$

for T = 10 K, where $k_B$ is Boltzmann constant. Even using this approximate value in Eq. 10, we then get the threshold density for 2D excitons as $N_{BEC}^{2D}$ ~$1.52 \times 10^{11}$/cm². This purely 2D estimate is clearly close to the calculated surface charge density $(\sigma_{ph})$ of photo generated indirect excitons at 10 K even for our quasi 2D structure as mentioned above. Moreover, this $\lambda_{Th}$ =100 nm at 10 K is certainly much bigger than the ~10 nm in-plane separation[7] between neighboring InAs QDs with surface density ~$10^{11}$/cm². Therefore, it is likely that actual BEC coherence length can be even larger and wave functions of these quantum coupled indirect excitons begin to overlap cooperatively with each other around 70 K. It is interesting to note here that without any in-plane localization of excitons using InAs QDs, such lateral overlap of bosonic wave functions won't be easy to accomplish in the standard coupled 2D-2D bi-layer structures generally used for studying excitonic BEC.

Therefore, from the observation of phase matched coherent oscillations, we can actually expect these indirect excitons to undergo BEC within a lateral span of ~200 μm in XY plane. In fact, we find the connection between critical density of BEC as a function of thermal de-Broglie wavelength using the following formula[41].

$$n_{EX} \sim N_{BEC}^{2D} = -\left(\frac{1}{\lambda_{Th}^2(T_C^{BEC})}\right) ln\left(1 - \exp(-\frac{|\epsilon_0|}{k_B T_C^{BEC}})\right) \tag{12}$$

In Fig. 5d, we plot this critical temperature ($T_C^{BEC}$) for attaining excitonic BEC in this 0D-2D system as function of surface density ($n_{EX}$) of indirect excitons with $|\epsilon_0| = 6.8$ meV as the approximate ground state binding energy[41,42] of these indirect excitons. With 2D photoexcitation intensity of nearly ~$3 \times 10^{16}$/cm² at 630 nm, we expect to generate a large number of indirect excitons as $n_{EX} \sim \sigma_{Ph} \sim 10^{11}$/cm² as shown above. From Fig. 5d, we clearly see that even at a nominal indirect exciton density of around one per QD as $n_{EX} \sim \sigma_{Ph} \sim 10^{11}$/cm², the expected transition temperature of excitonic BEC in two dimensions is certainly around 70 K or even higher. We also understand, that it is possible that our quasi 2D system may actually be undergoing a Berezinskii-Kosterlitz-Thouless (BKT) like phase transition where the quasi long-range order is only non-vanishing within a finite region. We also plot the critical temperature for a BKT like transition using the following formula[18,43] in the same plot where $a_0$ is the Bohr exciton radius.

$$k_B T_C^{BKT} = \frac{(\hbar^2/2m^*)4\pi n_{EX}}{\ln\left(\ln\left(\frac{1}{n_{EX}a_0^2}\right)\right)} \tag{13}$$

The estimated transition temperature for BKT transition is somewhat lower than the above estimate for 2D BEC but it is still around 70 K or more for $n_{EX} \sim \sigma_{Ph} \sim 10^{11}$/cm² even when we assume $a_0 \approx 10$ nm following[41,42] $|\epsilon_0| = 6.8$ meV. So, we can also expect a BKT like phase transition to develop over this large area.

Therefore, we now state that these 0D-2D indirect excitons are likely to form a large phase coherent quantum state of (say) $n_{EX}$ excitons covering ~200 μm areal diameter in XY plane below 70 K. Eventually, constituent electrons of these indirect excitons in BEC or BKT state can produce these nicely phase matched oscillations through coherent resonant tunneling.

So, our observations indicate an underlying many-body phase transition of these photo generated and bias driven, spatially indirect excitons. This is because the areal density of these condensed bosons (indirect excitons) as BEC order parameter is also the limiting value of first order spatial correlation function $g_1(r)$ such that[44,45]

$$n_{EX} \sim \sigma_{Ph} \equiv \lim_{r \to \infty}[g_1(r)] \qquad (14)$$

In our experiments we see this as $n_{EX} \sim \sigma_{Ph} \sim 10^{11}/cm^2$ even when $r \to 200\ \mu m$. As a result, here we again expect the interacting Bose gas of dipolar, indirect excitons to exhibit quasi long-range order over macroscopically large area having diameter ~200 μm. Although, we have not directly measured this off-diagonal long-range order parameter in terms of $g_1(r)$, which is currently beyond the scope of this study. However, it is understandable that any strong density fluctuations of these excitons over such a wide region could have easily destroyed any matching phase coherence between its constituent electrons undergoing resonant tunneling. Observation of coherent, almost sinusoidal oscillations, therefore, necessarily requires one to sustain the phase coherence of these excitons being probed with this large electrical mesa contact. Moreover, excitonic condensation can in principle drive its constituent electrons in the GaAs 2DEG to have a single unique momentum state and to have an excitonic polarization vector sharply aligned along $\vec{z}$ below a threshold temperature as discussed in the last section in connection with our observations shown in Fig. 5c. In fact, this sudden increase in the BEC order parameter in the form of surface charge density $n_{EX}$ or average electric dipolar polarization vector $\langle \vec{P} \rangle$ of these indirect excitons is measured by increased photocapacitance (Eq. 6 and Eq. 7). This can also be viewed as occurrence of some spontaneous symmetry breaking associated with excitonic BEC in quantum coupled 0D-2D system. A schematic quantum mechanical description of these dipolar indirect excitons before and after BEC is now shown in Fig. 6. Subsequently, we witness the development of a macroscopic quantum state of excitonic BEC and associated quasi long-range order over an areal diameter of ~200 μm. Moreover, excitonic condensation and related enhancement of electric polarization also allow for alignment of these indirect excitonic dipoles in a mutually parallel configuration along the growth axis (z axis). It is well known that similar parallel configurations of bosonic dipoles are also suitable for achieving a stable dipolar BEC[46]. However, the problem still remains open for future investigations of any spatio-temporal indicators of either BEC or BKT states of excitons in this quantum coupled 0D-2D heterostructure and also to know the exact fractions of dark[19] to bright excitons during the oscillations.

## IX. Additional supporting discussions on excitonic BEC.

Experimental signatures of these coherent oscillatory quantum phenomena are explored here with photocapacitance oscillations and with resonantly sharp excitonic spectral shapes at peak oscillation biases. These are observed when electrons coherently tunnel out of the GaAs TQW into InAs QDs. Presence of coherent Fabry-Perot like resonant interferometric tunneling of electrons at peak oscillation biases actually allows for sharp excitonic spectral shapes. This happens when the excitonic ensembles are driven towards excitonic BEC and quantum tunneling strongly couples[28] the 0D-2D system. These sharp excitonic spectral shapes at peak oscillation biases may also point towards the presence of a unique quantum coherent state of indirect excitons. Spectroscopic evidences of this resonantly sharp, doubly split, stationary states of excitons in this strongly quantum coupled heterostructure provides further support for coherent resonant tunneling as we still see photocapacitance oscillations even when the sample is photo excited with photons having much smaller energy than ~1.61 eV (Fig. 1d).

On the other hand, we understand that this coupled excitonic system coherently evolves into a mixture of negatively charged trions and indirect excitons at valley oscillation biases corresponding to decrease in tunneling current or minima of $dI_{Ph}/dV$ and negative[34] Photo-G/ω as shown in Fig. 1c. These change the dipolar environment of these indirect excitons considerably, which manifest itself through an uncharacteristically broad spectral shape of photocapacitance in Fig. 2b. We also observed that excitonic BEC and subsequent coherent resonant tunneling perpendicular to the heterojunction can be tuned with applied biases possibly in a way somewhat similar to that proposed earlier[39] as electric field induced Aharonov-Bohm effect for in-plane currents. Nevertheless, resultant quantum coherent oscillation between bosonic ensemble of excitons and fermionic ensemble of trions is thereby making itself observable as coherent photocapacitance and photo-G/ω oscillations because the spatial phase coherence is built up over a fairly large region with diameter ~200 μm.

In summary, we think that strong in-plane localization of indirect excitons having holes inside InAs QDs is one of the main reason for observing excitonic condensations in this heterostructure at temperatures below 70 K. Areal density[7] of these InAs QDs ~$10^{11}$/cm$^2$ is comparable to the estimated surface density of indirect excitonic dipoles as $n_{EX}$ ~$\sigma_{Ph}$ ~$10^{11}$/cm$^2$ under selective photo excitation intensity of ~$3\times10^{16}$/cm$^2$ at 630 nm using an incoherent lamp source. Therefore, it is likely that the number of distinct combinatory ways in which these indistinguishable bosons of indirect excitons can be accommodated in any available excited states of this coupled heterostructure is relatively smaller compared to the inverse of the Boltzmann factor[47]. As a result, the Boltzmann factor of order ~$e^{-n_{EX}}$ then drives this coupled 0D-2D excitonic system to its lowest energy BEC ground state. However, it is obvious that the distribution of these InAs QDs are also sufficiently dense so that the bosonic wave function of these coupled 0D-2D indirect excitons overlap strongly to produce a macroscopically large quantum state of excitonic BEC. This is already experimentally verified by significant population enhancement of excitonic ground state or BEC of electrically polarizable excitonic dipoles below 70 K (Fig. 5c) and also by the fact that the long-range phase coherence actually persists over a wide region covering ~200 μm.

Following Jan & Lee[41], we explain experimental observation of excitonic BEC to the close proximity of localized states in the form of InAs QDs near the GaAs 2DEG to spatially anchor these indirect excitons for optimal overlap of bosonic wave functions. This spatial confinement of these excitonic bosons in the XY plane not only provides a finite, discreet ground state energy level which the excitonic chemical potential can approach from below but also a dense enough in-plane distribution of these QDs to form this excitonic BEC. Anchoring of holes of these indirect excitons within InAs QDs further prevents them from coming too close to precipitate any unwanted Mott transitions at high densities which could have eventually destroyed any macroscopic phase coherence of the BEC quantum state. In fact, such ideal phase matching is eventually lost through decoherence as the quantum oscillation magnitudes slowly vanish with increasing DC bias currents.

Moreover, we understand that repulsive dipolar interaction of these localized indirect excitons within this quantum coupled 0D-2D heterostructure can also be helpful in stabilizing excitonic BEC with quasi long-range order spanning a wide region as repulsive interactions usually reduce long-range density fluctuations[48]. We speculate that this BEC state of indirect excitons with holes localized inside these QDs can be further stabilized[46] through parallel configuration of dipole-dipole repulsive interactions in a more densely packed quasi 2D, coupled quantum dot-quantum well heterostructures.

## X. Conclusions.

We observed coherent photocapacitance oscillations of excitonic states in coupled 0D-2D heterostructure, which establish the presence of macroscopically large quantum state of excitonic BEC. Such BEC is sensed through observation of phase coherent resonant tunneling, which requires many-body cooperative effects over a wide region ~200 μm. Estimated thermal de-Broglie wavelength is nearly an order of magnitude larger than in-plane separation between InAs QDs around 10 K. Consequently, this produces a density driven phase transition of average electric polarization vector of these excitonic dipoles and generate excitonic BEC below 70 K. Formation of this many-body, quantum coherent state was essential to produce the necessary cooperative effects through Fabry-Perot type resonant tunneling to sustain the photocapacitance oscillations.

We elaborate on how this macroscopically large quantum state of excitonic BEC can be controlled with electrical biases using this coupled system of quantum dots (0D) and quantum well (2D). Attaining such optimum localized configuration and critical density of excitonic bosons with overlapping excitonic wave functions were certainly more difficult in the standard, planar 2D-2D bi-layer samples. Therefore, we also comprehend why this excitonic BEC was so difficult to achieve earlier in any such coupled quantum well samples every so often employed for experimental investigations of BEC of excitons till now.

We also report interesting spectroscopic observations of coherent, phase matched oscillations between BEC of spatially indirect excitons and fermionic mixture of charged trions. Finally, we also speculate why such interacting dipolar systems of spatially localized indirect excitons in similar quantum coupled 0D-2D heterostructures can be a promising candidate for observing electrically tunable excitonic BEC even around room temperatures using materials having larger excitonic binding energies.

## XI. Samples and Experimental Methods.

The sample used for our measurements is exactly similar to one of those used by Vodovin et al[7,8,9]. It has one InAs quantum dots layer within single barrier GaAs/AlAs/InAs/AlAs/GaAs p-i-n diode. It was grown by Molecular Beam Epitaxy (MBE) on a highly doped $n^+$ GaAs substrate. The following layers are grown subsequently on the substrate: a 1.0 μm heavily doped ($4\times10^{18}$ cm$^{-3}$) $n^+$ GaAs layer is grown at temperature 550 ºC on top of the substrate for bottom contact, subsequently a 100 nm n-doped ($2\times10^{16}$ cm$^{-3}$) GaAs layer, and then 100 nm undoped spacer layer of GaAs, two intrinsic AlAs layers of thickness 5.1 nm each having a 1.8 monolayer of InAs quantum dots layer in between them, then a 60 nm undoped GaAs layer complete the intrinsic region, after that a 0.51 μm heavily doped ($2\times10^{18}$ cm$^{-3}$) $p^+$ GaAs layer which facilitates top Ohmic contact for the device. Net area of the ring-shaped gold contact pads is approximately $3\times10^{-4}$ cm$^{-2}$.

For low temperature measurements, we placed our sample on a customized copper holder inside a closed-cycle cryostat CS-204S-DMX-20 from Advance Research Systems. The temperature of the cryostat is controlled with a Lakeshore (Model-340) temperature controller.

For photocapacitance measurements ($C_{photo}$= dQ/dV), the sample is illuminated from the top p-GaAs side using an Acton Research SP2555 monochromator having a 0.5-m focal length along with a 1000-W quartz-tungsten-halogen lamp as non-coherent light source. For photocapacitance measurements, we used Agilent's E4980A LCR meter with small signal RMS AC voltage of 30 mV with a frequency of 10 kHz (unless mentioned otherwise). A simple series equivalent circuit of capacitance (C) and conductance (G) in parallel was used to extract these parameters. The spectral response of the lamp-monochromator combination is reasonably smooth and changes slowly and monotonically within the wavelength ranges we use in our experiments. Lambda square corrections[49] were not used while plotting these spectra.

## XII. ACKNOWLEDGEMENTS

SD acknowledges Department of Science and Technology (DST), India (Grants # DIA/2018/000029 and SR/NM/TP13/2016). AB and MKS are thankful to DST, India for Inspire PhD Fellowship and IISER-Pune for Integrated PhD Fellowship, respectively. AB also received support from the Newton-Bhabha Ph.D Placement Programme of DST, India and British Council, UK. MH acknowledges support from the UK Engineering and Physical Sciences Research Council. M.H. and MAH are grateful for the support by a grant from the deanship of scientific research, Princess Nourah bint Abdulrahman University, Riyadh, KSA.

**Table 1. Theoretical and experimental comparison of blue shifts of photocapacitance spectra under applied bias.**

| Crest Bias (V) | TQW Energy level (n=1) (meV) | Lowest peak energy from the excitonic spectra (eV) | Spectral Energy difference for successive peak oscillation biases (meV) | Estimated change in TQW energy level (n=1) for successive peak oscillation biases (meV) |
|---|---|---|---|---|
| $V_1$= -0.4 | 94.8 | 1.520 | 8.4 (for $V_2,V_1$) | 8 (for $V_2,V_1$) |
| $V_2$= -0.66 | 103.2 | 1.528 | 8.2 (for $V_3,V_2$) | 9 (for $V_3,V_2$) |
| $V_3$= -0.92 | 111.4 | 1.537 | 16.6 (for $V_3,V_1$) | 17 (for $V_3,V_1$) |

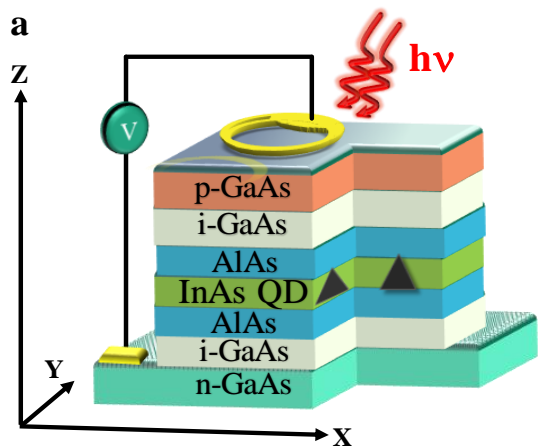
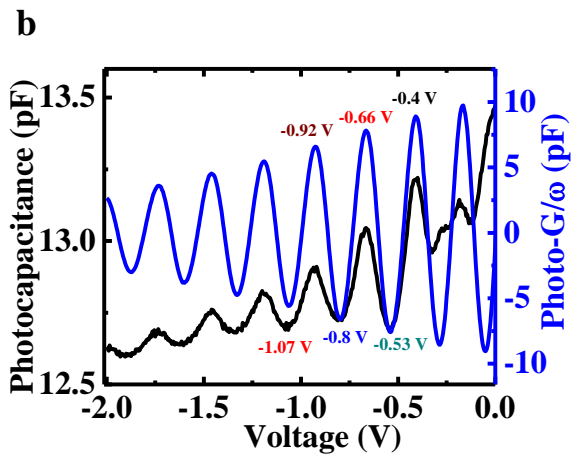
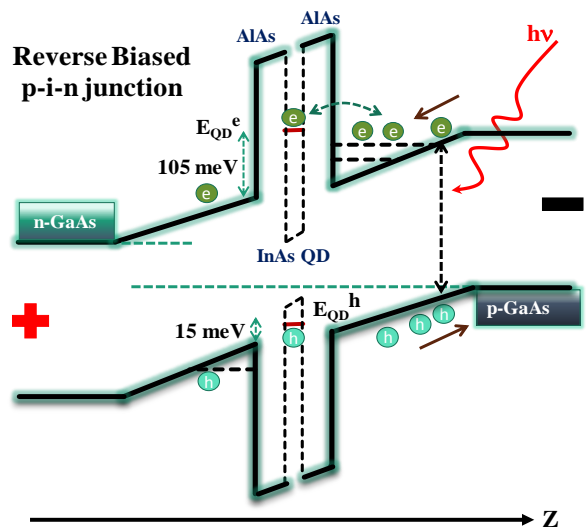
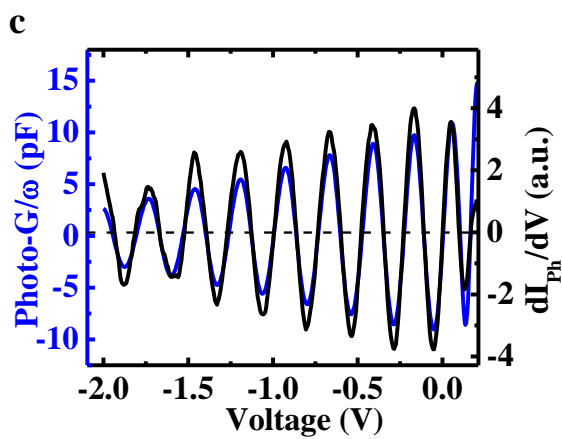
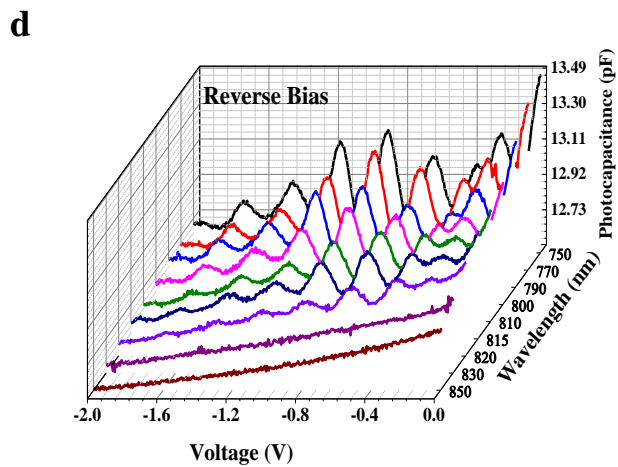
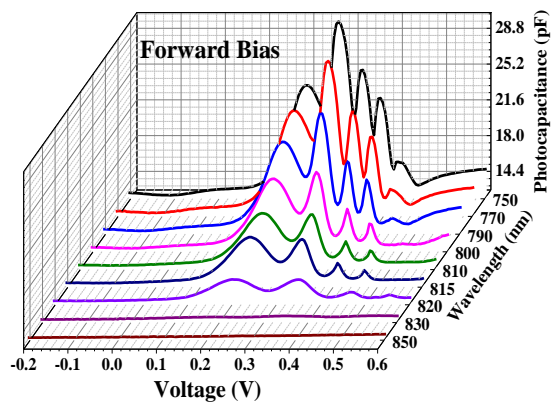

**Figure 1:** Coherent oscillations in photocapacitance and photo conductance at 10 K. (a) Schematic diagram of the GaAs/AlAs/InAs/AlAs/GaAs sample and corresponding energy band-diagram under reverse bias are depicted. The '+' and '-' signs indicate the direction of applied reverse bias. Brown arrows indicate the drift direction of electrons and holes under reverse bias. Growth direction $\hat{Z}$ for both sample and band diagram are drawn separately. (b) Quantum oscillations of photocapacitance and Photo-G/ω with the sample under reverse bias are shown under selective photoexcitation with 630 nm. (c) Shows how the photo-G/ω and $dI_{Ph}/dV$ oscillations coincide, $I_{Ph}$ being the DC-photo current. (d) Photocapacitance oscillation under selective photo excitations in both reverse and forward bias are shown separately. These oscillations survive even when photo excited below ~1.61 eV but fully subside for photo excitations below GaAs bandgap. 3D plots are slightly tilted for better visibility.

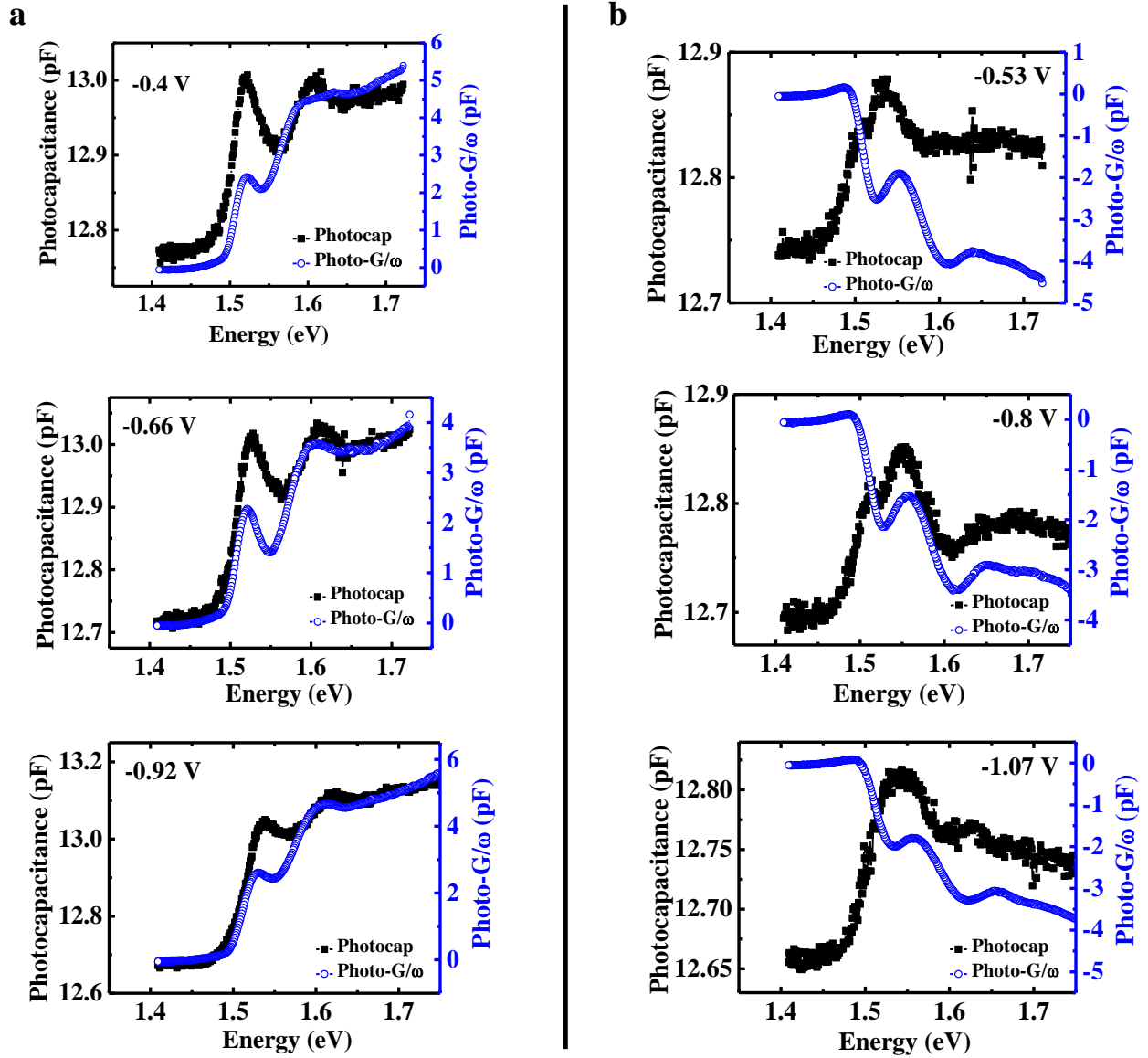

**Figure 2:** Presence and absence in two excitonic transitions in photocapacitance spectra at peak and valley oscillation biases at 10 K. (a) Photocapacitance and photo-G/ω spectra under biases of successive peaks of photocapacitance oscillations from Fig. 1b. Corresponding biases are -0.40 V, -0.66 V, -0.92 V. We see the presence of two excitonic features. (b) Photocapacitance and photo-G/ω spectra under biases of successive valleys of photocapacitance oscillations from Fig. 1b. Corresponding biases are -0.53 V, -0.80 V, -1.07 V. Those two sharp excitonic states from the plots in Fig. 2a on the left panel are strikingly missing from the photocapacitance spectra in Fig. 2b on the right panel.

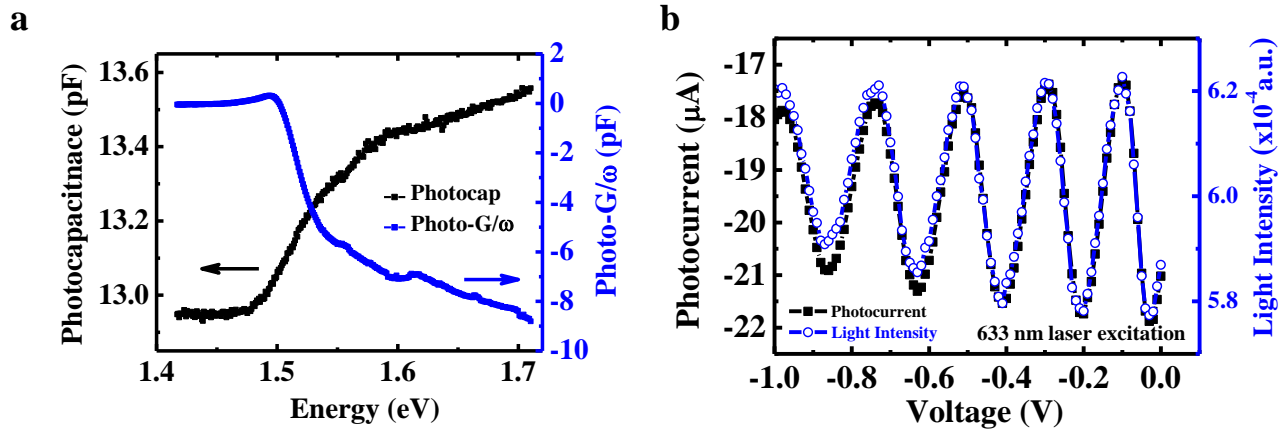

**Figure 3:** Absence of excitonic spectra at zero bias and photoluminescence oscillations at 10 K. (a) Excitonic peaks of both GaAs and InAs QD are hardly visible under zero bias which indicates the critical roles of applied bias in these measurements. (b) Photocurrent and integrated photoluminescence intensity in the front collection geometry oscillate similarly with increasing reverse bias. Both of these are measured using 633 nm He-Ne laser excitation.

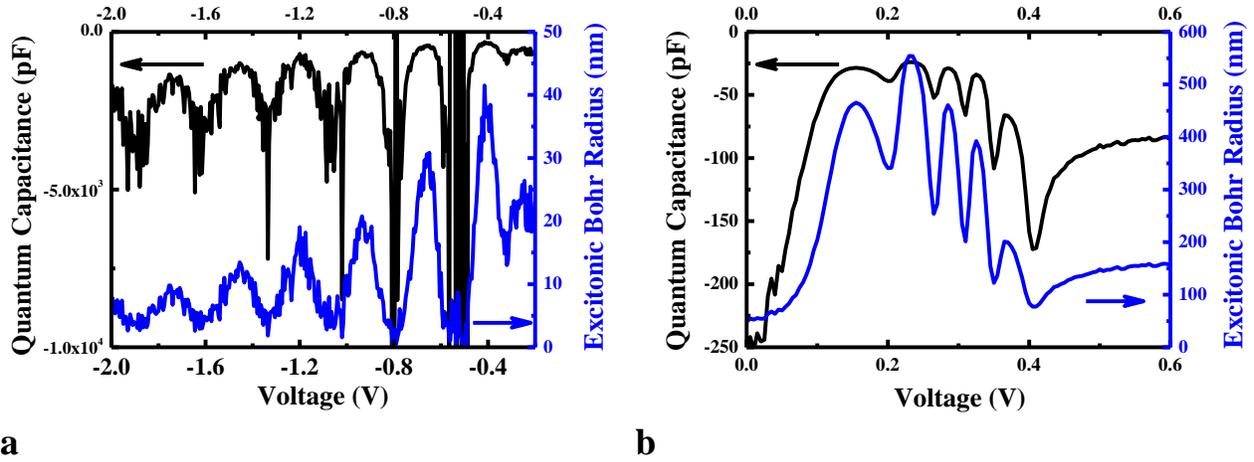

**Figure 4:** Excitonic Quantum Capacitance in reverse bias and corresponding Bohr radius at 10 K. Oscillations of quantum photo capacitance and Bohr radius ($a_0^*$) under (a) reverse bias (b) forward bias. Variation of quantum capacitance and $a_0^*$ in units of nm at successive peak oscillation biases are estimated from photocapacitance spectra in Fig. 2a. Average dipolar separation of these excitons scales as ($a_0^*/4$) under different reverse biases.

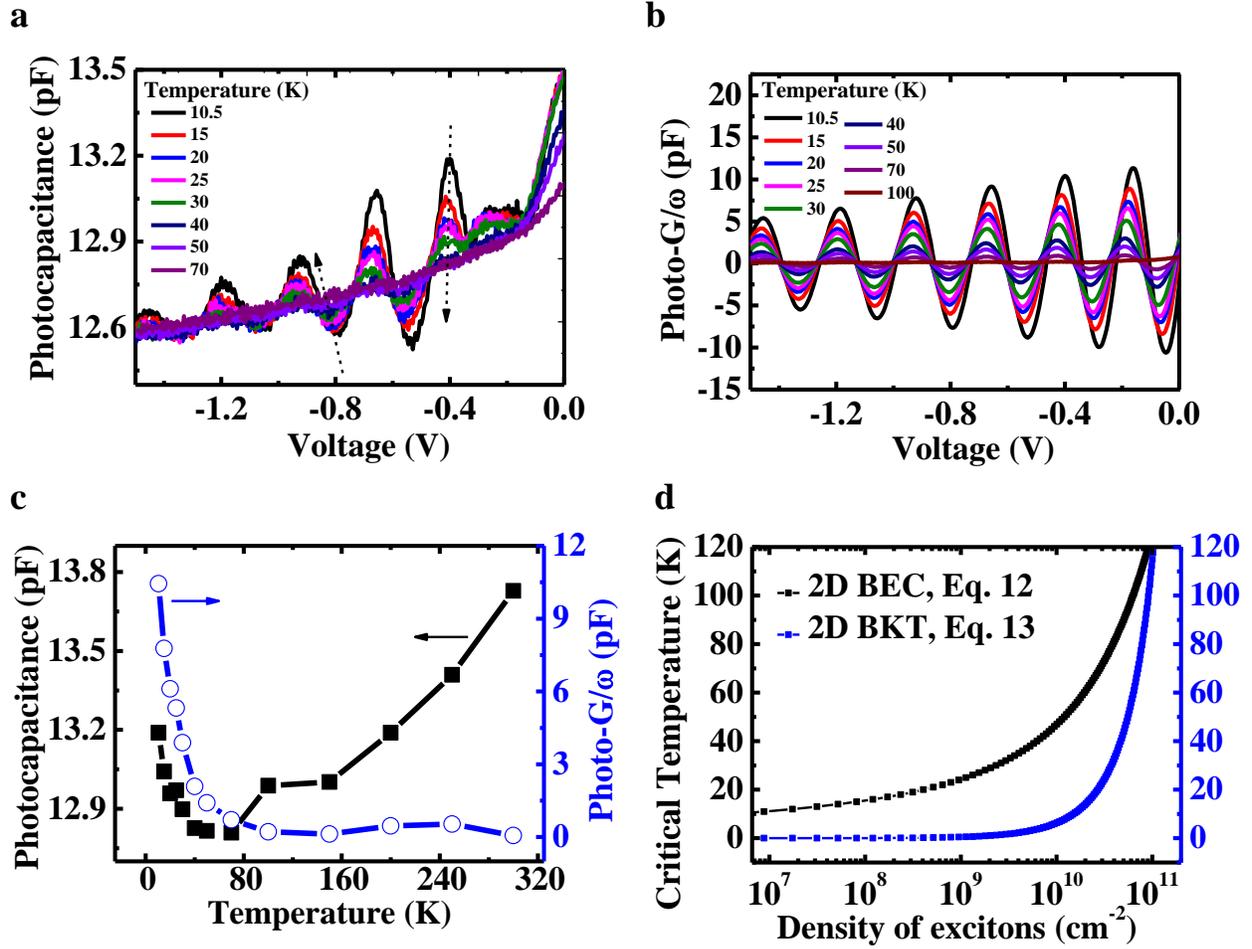

**Figure 5:** Temperature dependence of photocapacitance oscillations and possible phase transition. (a) Photocapacitance oscillations at different temperatures. (b) Photo-G/ω oscillations at different temperatures. (c) Variation of photocapacitance and Photo-G/ω at -0.40 V reverse bias as a function of temperature. It is clear that surface charge density and associated electric polarization of these 0D-2D excitonic dipoles suddenly increase below ~70 K, indicating a density-driven phase transition. (d) Simulated plots of critical temperature of 2D BEC and 2D BKT transitions as a function of 2D density of excitons following Jan & Lee[41] and Snoke[18], respectively.

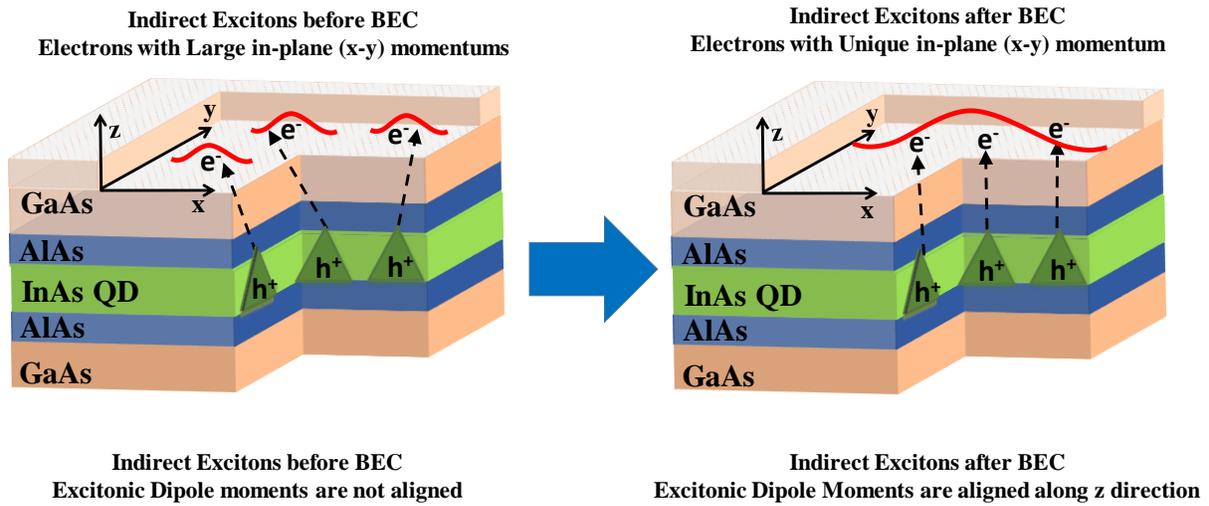

**Figure 6:** Schematic diagram of interacting dipoles of indirect excitons undergoing momentum space narrowing after BEC transition in this quantum coupled 0D-2D heterostructure. The red contours on the left represent the wave functions of individual indirect excitons, whereas the right one represents the macroscopic quantum state of excitonic BEC. This is because the thermal de-Broglie wavelength (~100 nm at 10 K) of these indirect excitons easily exceeds the average 10 nm spacing between the neighboring InAs QDs. Spatial anchoring of these indirect excitons with InAs QDs is crucial to form the BEC state.